\documentclass[9pt,conference,a4paper]{IEEEtran}
\ifCLASSINFOpdf
  \usepackage[pdftex]{graphicx}
\else
\fi
\usepackage{subfig} 
\usepackage{float}

\def\qunit{\mathbf{\hat{q}}}
\newcommand{\figref}[1]{Fig.\,\ref{#1}}
\begin{document}

\title{Multi-shell Sampling Scheme with Accurate and Efficient Transforms for Diffusion MRI}

\author{%
\IEEEauthorblockN{
Alice P. Bates\IEEEauthorrefmark{1},
Zubair Khalid\IEEEauthorrefmark{1},
Rodney A. Kennedy\IEEEauthorrefmark{1}
and Jason D. McEwen\IEEEauthorrefmark{2}}
\IEEEauthorblockA{\IEEEauthorrefmark{1}
Research School of Engineering,
Australian National University,
ACT 2601,
Australia.}
\IEEEauthorblockA{\IEEEauthorrefmark{2}
Mullard Space Science Laboratory,
University College London,
Surrey RH5 6NT,
UK.}}

\maketitle

\begin{abstract}
We propose a multi-shell sampling grid and develop corresponding transforms for the accurate reconstruction of the diffusion signal in diffusion MRI by expansion in the spherical polar Fourier (SPF) basis. The transform is exact in the radial direction and accurate, on the order of machine precision, in the angular direction. The sampling scheme uses an optimal number of samples equal to the degrees of freedom of the diffusion signal in the SPF domain.


\end{abstract}

\section{Introduction}
The diffusion signal in diffusion MRI can be reconstructed from a finite number of measurements in $q$-space, where $\mathbf{q}$ is the diffusion wavevector, from which the brain's connectivity and microstructure properties can be determined. In diffusion MRI, the number of measurements that can be acquired is highly restricted due to the need for scan times to be practical in a clinical setting. For this reason, multi-shell sampling schemes, where samples are collected on multiple concentric spheres with different $q$-space radii, are commonly used rather than large Cartesian sampling grids. Existing multi-shell sampling schemes require more than the optimal number of samples, defined as the degrees of freedom in the basis used to reconstruct the diffusion signal, in order to allow for the accurate reconstruction of the diffusion signal and use least-squares to calculate coefficients, which is computationally intensive (e.g. \cite{assemlal:2009b}).

The spherical polar Fourier (SPF) basis \cite{assemlal:2009b} is a 3D complete orthonormal basis commonly used for reconstructing the diffusion signal. The normalised MR signal attenuation, $E(\mathbf{q})$ can be expanded in the SPF basis, as
\begin{equation}
\label{Eq:E_expansion}
    E(\mathbf{q})=\sum_{n=0}^{N-1}\sum_{\ell=0}^{L-1}\sum_{m=-{\ell}}^{\ell} E_{n\ell m}  R_n(q)Y_{\ell}^m(\qunit),
\end{equation}
where $\qunit = \frac{\mathbf{q}}{|\mathbf{q}|}$, $q = |\mathbf{q}|$,  $Y_{\ell}^m(\qunit)$ are spherical harmonic coefficients of degree $\ell$ and order $m$, and the expansion coefficients are given by
\begin{equation}\label{Eq:Ecoeff}
E_{n\ell m} = \langle E(\mathbf{q}),  R_n(q)Y_{\ell}^m(\qunit)\rangle.
\end{equation}
The radial functions are Gaussian-Laguerre polynomials $R_n$ with
\begin{equation}
\label{Eq:radial_func}
    R_n(q)=\bigg[\frac{2}{\zeta^{1.5}}\frac{n!}{\Gamma(n+1.5)}\bigg]^{0.5}\exp\bigg(\frac{-q^2}{2\zeta}\bigg)L_n^{1/2}\bigg(\frac{q^2}{\zeta}\bigg),
\end{equation}
where $\zeta$ denotes the scale factor and $L_n^{1/2}$ are the $n$-th generalised Laguerre polynomials of order half. The expansion Eq. (\ref{Eq:E_expansion}) assumes that $E(\mathbf{q})$ is band-limited at radial order $N$ and angular order $L$.

\section{Multi-shell Sampling Scheme and SPF Transform}
The 3D transform for calculating the diffusion signal coefficient (Eq.\ref{Eq:Ecoeff}) can be separated into transforms in the radial and angular directions due to the separability of the SPF basis. For the radial transformation, Gauss-Laguerre quadrature can be used, where $N$ sampling nodes is sufficient for exact quadrature. The $N$ shells of our proposed multi-shell sampling scheme are placed at $q_i = \sqrt{\zeta x_i}$ where $x_i$ are the roots of the $N$-th generalised Laguerre polynomial of order a half. We determine the corresponding weights to be
\begin{equation}
\label{Eq:weight}
w_i =\frac{0.5\zeta^{1.5}\Gamma(N+1.5)x_ie^{x_i}}{N!(N+1)^2[L_{N+1}^{0.5}(x_i)]^2}.
\end{equation}
The number of shells required for accurate reconstruction was recommended as $N=4$ in~\cite{assemlal:2009}.  We set the scaling factor $\zeta$ so that shells are located at $b$-values within an interval of interest. In this work, we use a maximum $b$-value of 8000 $\rm{s/mm}^2$, resulting in shells at $b = 411.3, 1694.4, 4036.3$ and $8000$ $\rm{s/mm}^2$.

\begin{figure}[t]
  \centering
  \vspace{-5mm}
  \subfloat[]{
  \includegraphics[width=0.22\textwidth]{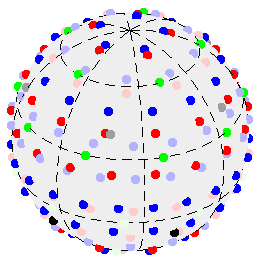}}
   \subfloat[]{
  \includegraphics[width=0.22\textwidth]{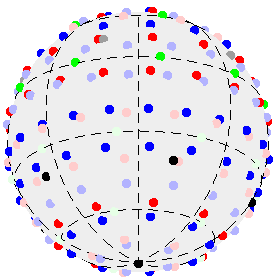}}
  \vspace{-2mm}
  \caption{(a) North pole view and (b) South pole view.}
  \label{fig:sampling_grid}
  \vspace{-7mm}
\end{figure}

For sampling within each shell, we use the recently proposed single-shell sampling scheme~\cite{Bates:2015} which allows accurate reconstruction on the order of machine precision accuracy, has an efficient forward and inverse spherical harmonic transforms, and uses an optimal number of samples for the band-limited diffusion signal on the sphere, $L(L+1)/2$. The spherical harmonic band-limit, and therefore the number of samples within each shell, is determined using \cite{daducci:2011}, where the authors determined the spherical harmonic band-limit $L$ required to accurately reconstruct the diffusion signal at different b-values, the shells have $L =$ 3, 5, 9 and 11 respectively. The proposed sampling scheme has a total of 132 samples. \figref{fig:sampling_grid} shows the sampling scheme projected onto a single sphere, samples are shown in black, green, red and blue for each shell respectively. Locations where antipodal symmetry is used to infer the value of the signal are lighter in color. 

%

\vspace{-4mm}
\bibliographystyle{IEEEtran}
\bibliography{bib}

\begin{thebibliography}{1}
\providecommand{\url}[1]{#1}
\csname url@samestyle\endcsname
\providecommand{\newblock}{\relax}
\providecommand{\bibinfo}[2]{#2}
\providecommand{\BIBentrySTDinterwordspacing}{\spaceskip=0pt\relax}
\providecommand{\BIBentryALTinterwordstretchfactor}{4}
\providecommand{\BIBentryALTinterwordspacing}{\spaceskip=\fontdimen2\font plus
\BIBentryALTinterwordstretchfactor\fontdimen3\font minus
  \fontdimen4\font\relax}
\providecommand{\BIBforeignlanguage}[2]{{%
\expandafter\ifx\csname l@#1\endcsname\relax
\typeout{** WARNING: IEEEtran.bst: No hyphenation pattern has been}%
\typeout{** loaded for the language `#1'. Using the pattern for}%
\typeout{** the default language instead.}%
\else
\language=\csname l@#1\endcsname
\fi
#2}}
\providecommand{\BIBdecl}{\relax}
\BIBdecl

\bibitem{assemlal:2009b}
H.~Assemlal, D.~Tschumperl{\'e}, and L.~Brun, ``Efficient and robust
  computation of {PDF} features from diffusion {MR} signal,'' \emph{Med. Image
  Anal.}, vol.~13, pp. 715--729, Jun. 2009.

\bibitem{assemlal:2009}
H.~Assemlal, D.~Tschumperl\'{e}, and L.~Brun, ``Evaluation of q-space sampling
  strategies for the diffusion magnetic resonance imaging,'' in \emph{Med.
  Image Comput. Comput. Assist. Interv., MICCAI'2009}, London, UK, 2009,
  vol.~12, no. Pt 2, pp. 406--414.

\bibitem{Bates:2015}
A.~P. Bates, Z.~Khalid, and R.~A. Kennedy, ``An optimal dimensionality sampling
  scheme on the sphere for antipodal signals in diffusion magnetic resonance
  imaging,'' in \emph{Proc. IEEE Int. Conf. Acoust., Speech, Signal Process.,
  ICASSP'2015}, Brisbane, Australia, Apr. 2015, pp. 872--876.

\bibitem{daducci:2011}
A.~Daducci, J.~D. McEwen, D.~V.~D. Ville, J.~P. Thiran, and Y.~Wiaux,
  ``Harmonic analysis of spherical sampling in diffusion {MRI},'' in
  \emph{Proc. 19th Ann. Meet. Int. Soc. Magn. Reson. Med.}, Jun. 2011.

\end{thebibliography}

\end{document}